\documentclass[journal]{IEEEtran}

\ifCLASSINFOpdf
\else
  \usepackage[dvips]{graphicx}
  \graphicspath{{../eps/}}
 \fi
\begin{document}
\title{Speckles-Training-Based Denoising Convolutional Neural Network Ghost Imaging}

\author{Yuchen~He,
        Sihong~Duan,
        Jianxing~Li,
        Hui~Chen,
        Huaibin Zheng,
        Jianbin Liu,
        Shitao~Zhu,
        and~Zhuo~Xu
\thanks{Manuscript received Apr 7, 2021. This work was supported in part by the National Natural Science Foundation of China (Grant No. 61901353), in part by the Fundamental Research Funds for the Central Universities (No. xjh012019029), in part by the 111 Project of China (Grant No. B14040).(Corresponding authors: Jianxing Li; Hui Chen.)}
\thanks{Y. He, H. Chen, H. Zheng, J. Liu, Z. Xu, are with School of Electronic Science and Engineering, Xi'an Jiaotong University, Xi'an 710049, China. (email: yuchenhe@xjtu.edu.cn, chenhui@xjtu.edu.cn, huaibinzheng@xjtu.edu.cn, liujianbin@xjtu.edu.cn, xuzhuo@xjtu.edu.cn)}
\thanks{J. Li, S. Zhu, are with School of Information and Communications Engineering, Xi'an Jiaotong University, Xi'an 710049, China. (email: jianxingli.china@xjtu.edu.cn, shitaozhu@xjtu.edu.cn).}
\thanks{S. Duan, is with School of Software Engineering, Xi'an Jiaotong University, Xi'an 710049, China. (email: duansihong@stu.xjtu.edu.cn).}}


\maketitle

\begin{abstract}
Ghost imaging (GI) has been paid attention gradually because of its lens-less imaging capability, turbulence-free imaging and high detection sensitivity.
However, low image quality and slow imaging speed restrict the application process of GI.
In this paper, we propose a improved GI method based on Denoising Convolutional Neural Networks (DnCNN).
Inspired by the corresponding between input (noisy image) and output (residual image) in DnCNN, we construct the mapping between speckles sequence and the corresponding noise distribution in GI through training.
Then, the same speckles sequence is employed to illuminate unknown targets, and a de-noising target image will be obtained.
The proposed method can be regarded as a general method for GI.
Under two sampling rates, extensive experiments are carried out to compare with traditional GI method (basic correlation and compressed sensing) and DnCNN method on three data sets.
Moreover, we set up a physical GI experiment system to verify the proposed method.
The results show that the proposed method achieves promising performance.
\end{abstract}


\IEEEpeerreviewmaketitle

\section{Introduction}
\IEEEPARstart{G}{host} imaging (GI) exploits second-order correlation for imaging~\cite{pittman1995optical}, making it quite different than conventional methods based on point to point (or spot) scheme.
GI illuminates target with a sequence of light patterns and measures the total light intensity reflected (or transmitted) from the target with a bucket detector.
Target image is then reconstructed by calculating the correlation between illumination patterns and corresponding bucket detections.
Because of lens-less imaging capability, turbulence-free imaging and high detection sensitivity, GI has invoked a large body of literature during the past two decades, more and more researchers have been seeking its applications~\cite{shapiro2008computational, meyers2011turbulence, pelliccia2016experimental, khakimov2016ghost, Ota1246}.
On the other hand, this scheme has been suffering from slow imaging speed and low image quality, hindering the development of its applications.

Sampling rate (the number of illumination patterns dividing the number of resolution pixels) primarily determines the imaging speed in mathematics.
Low sampling rate will accelerate imaging but degrade the quality especially when high resolution is demanded.
This conflict is a barrier to the application of GI.
During the past decade, researchers have shown an increased interest in solving this problem.
Orthogonal bases such as Hadamard, Fourier or Wavelet patterns can provide much better image quality than that of using random patterns. However, much more advanced technologies are required to implement the illumination when imaging speed is concerned~\cite{xi2019bi}.
Compressed sensing (CS) has been adopted to reduce the sampling rates of GI, but it usually hard to balance image quality and imaging speed~\cite{katz2009compressive}.
From 2017, artificial intelligence methods have been applied to GI and improved the imaging quality in certain situations~\cite{shimobaba2018computational, he2018ghost, wang2018de}.
Nevertheless, most of them merely perform a post-process on the images already recovered by a GI system, rather than integrate the experimental parameters (such as the bases) in the network.

Besides a low sampling rate, a fast imaging speed of GI requires illuminating patterns at a high speed.
In practical, it is not easy to illuminate a set of orthogonal bases at a high speed. Recently, a ultra-fast GI technique has been proposed, which used a LED array to emit Hadamard patterns (or any preset patterns) at 100 MHz rate, but the resolution is very limited~\cite{zhao2019ultrahigh}.
When the resolution increases, the illumination speed will decrease, and it also increase the cost.
There is another practical way to implement a fast illumination for random patterns: by manipulating the phase of a few light emitters, the interference of the emitted light-waves generates a random-like pattern.
The technique is of low cost and can perform high-resolution detection without raising the number of the emitters or reducing the modulation speed of the emission.
However, this technique cannot generate an arbitrary pattern as we want.
Usually, it can only generate a certain patterns that are highly subject to the number of the emitters and their spatial arrangement.
Moreover, the generated patterns are not well linearly independent, which will worsen the reconstructed image (usually containing ghost images).
A clear image can be post retrieved with a proper reconstruction algorithm, but the approach is not of universality.

For the purpose of application, GI should possesses a fast imaging speed, a good imaging quality, and a easy and inexpensive way to implement the detection. The previous researches have never integrated these three key factors in a study.
We here propose a framework based on Denoising Convolutional Neural Networks (DnCNN), which is aimed to seek an optimal imaging quality under a low sampling rate.
Importantly, the framework integrate the expansion bases used for a realistic measurement, which helps to reconstruct the best image for a practical GI system.
Recently, many results have been made in image denoising~\cite{xu2018external, zhang2018ffdnet, kokkinos2019iterative, kumar2019tchebichef, yao2019deep, frosio2019statistical, xu2020noisy, hou2020nlh, yang2020image, helou2020blind, gu2021blur}.
DnCNN was born in 2017 and is used for image denoising or super-resolution~\cite{zhang2017beyond}.
Compared with conventional denoising network, DnCNN uses residual learning method and adds batch normalization operation.
For a specific Gaussian noise level, DnCNN can reach the level of state-of-the-art (SOTA) in visual effect and numerical value.
DnCNN is trained by noise, the input is noisy image and the output is residual image.
Inspired by the corresponding between input and output in DnCNN, we use DnCNN to train a mapping between preset speckles and noises in GI.

In this paper, we report a improved GI method based on DnCNN, which learns the noise distribution of a set of speckles sequence, and use this speckles sequence to illuminate unknown targets.
Then, improved target image information can be obtained through the trained DnCNN.
For GI system, a sequence of preset speckles are employed to illuminate the imaging scene, and also utilized to reconstruct target image.
Regardless of target, there is a corresponding between noises and speckles.
Consequently, the proposed method can be considered as a general denoising method for GI system or even incoherent system, and denoise any target that is not in training set when same sequence speckles are used for training and illumination.

To verify the proposed method, we carry out a series of experiments.
Firstly, we compare the proposed method with the conventional methods in GI, BC method and CS method.
Then, we use the same training set to train a DnCNN in conventional way, and compare the proposed method and DnCNN at different $\sigma$ levels.
Further, to verify the corresponding relationship between denoising ability and speckles sequence, we use two sets of different speckle sequences to compare.
One of these two sets is utilized to train DnCNN, and we test these two sets of different speckle sequences on the trained DnCNN.
We carry out the above experiments on three different datasets, two different sampling rates, and the testing images in these datasets have not appeared in the training set.
Finally, we use actual random speckles sequence generated by spatial light modulator (SLM) to train DnCNN, and compare the proposed method with other methods.
The above experimental results show that the proposed method in this paper can achieve the best performance in various situations, and can be regarded as a general method for GI.

The main contributions of this work can be summarized as follows:

1) In this paper, we propose a improved GI method based on DnCNN.
Inspiring by DnCNN that train a mapping between noisy image and residual image, we construct the relationship between speckles and noises in GI system.
Therefore, we can illuminate target employing the speckles sequence that used in training to obtain better imaging result.

2) Extensive experiments show that the proposed method achieves promising performance on three different testing data sets under two sampling rates, especially in physical experiment.
Compared with other traditional methods in GI (BC and CS), or DnCNN, the proposed method has better performance in peak signal-to-noise ratio (PSNR) and structural similarity (SSIM).

3) The proposed method can be regarded as a general denoising method for GI.

The rest of this paper is organized as follows.
Section \uppercase\expandafter{\romannumeral2} provides a brief survey of related work.
In Section \uppercase\expandafter{\romannumeral3}, a comprehensive introduction to the proposed method is provided.
In Section \uppercase\expandafter{\romannumeral4}, the proposed method is compared with other methods by extensive experiments.
The paper is concluded in Section \uppercase\expandafter{\romannumeral5}.

\section{Related Work}
\subsection{Ghost Imaging}
From the view of mathematics, the process of GI is an optical expansion of the object in a set of spatial basis functions: each pattern represents a basis function, and the bucket detection is finding the corresponding coefficient. The correlation calculation is to reproduce the object by combining the bases and the coefficients together. In short, GI is an optical tool to acquire the spatial spectra of an object in the representation of certain bases. The selection of bases, how well to measure the coefficients and image reconstruction algorithm determine the imaging quality.

A typical GI system projects a set of light patterns (denoted as $\{P_1(\mathbf{r}),P_2(\mathbf{r}),...P_M(\mathbf{r})\}$) onto an object, where $M$ is the total number of the patterns. The illumination light is modulated by the object, and then reach the bucket detector that reads out the corresponding intensities as $\{b_1,b_2,...,b_M\}$. A bucket detection can be formulated as
\begin{equation}
b_j=\sum_{\mathbf r}P_j(\mathbf{r})O(\mathbf{r}),
\end{equation}
which is the $j$-th expansion coefficient on the $j$-th basic fucntion $P_j(\mathbf{r})$. The whole detection process can be  expressed in matrix form:
\begin{equation}
\left[ {\begin{array}{*{20}{c}}
{{b_1}}\\
{{b_2}}\\
 \vdots \\
{{b_M}}
\end{array}} \right] = \left[ {\begin{array}{*{20}{c}}
{{P_1}\left( {{{\bf{r}}_1}} \right)}&{{P_1}\left( {{{\bf{r}}_2}} \right)}& \cdots &{{P_1}\left( {{{\bf{r}}_N}} \right)}\\
{{P_2}\left( {{{\bf{r}}_1}} \right)}&{{P_2}\left( {{{\bf{r}}_2}} \right)}& \cdots &{{P_2}\left( {{{\bf{r}}_N}} \right)}\\
 \vdots & \vdots & \vdots & \vdots \\
{{P_M}\left( {{{\bf{r}}_1}} \right)}&{{P_M}\left( {{{\bf{r}}_2}} \right)}& \cdots &{{P_M}\left( {{{\bf{r}}_N}} \right)}
\end{array}} \right] \cdot \left[ {\begin{array}{*{20}{c}}
{O\left( {{{\bf{r}}_1}} \right)}\\
{O\left( {{{\bf{r}}_2}} \right)}\\
 \vdots \\
{O\left( {{{\bf{r}}_N}} \right)}
\end{array}} \right],
\end{equation}
or
\begin{equation}
\mathbf{B}=\mathbf{P}\cdot \mathbf{O}.
\end{equation}

\subsection{Denoising Convolutional Neural Networks}
The input of DnCNN is noisy image, which composed of original image and residual image. The output of DnCNN is residual image. The original image is eliminated in the process of training. The optimization target of DnCNN is not the mean-square error (MSE) between noisy image and original image, but the MSE between predicted residual image and real residual image. The loss function of DnCNN can be expressed as
\begin{equation}
\ell \left( \Theta  \right) = \frac{1}{{2N}}\sum\limits_{i = 1}^N {\left\| {\Re \left( {{y_i};\Theta } \right) - \left( {{y_i} - {x_i}} \right)} \right\|_F^2},
\end{equation}
where $\Re \left( {} \right)$ denotes the residual mapping $\Re \left( {{y_i}} \right) = {v_i}$. The residual learning formulation is utilized to train a residual mapping. ${y_i}$ is noisy image, ${v_i}$ is residual image. Then, we have ${x_i} = {y_i} - \Re \left( {{y_i}} \right)$, ${x_i}$ is the original image, and $\Theta$ is the trainable parameters, respectively. According to the principle of residual network, when the residual is 0, identity mapping is established, which is easy to train and optimize. During testing, high quality target image can be obtained quickly by subtracting prediction residual image from noisy image. DnCNN has a good denoising effect on a specific Gaussian noise level.

Inspired by the denoising effect of DnCNN on Gaussian noise, we consider introducing DnCNN into GI, and design training method that conform to the characteristic of GI. In GI, speckles are utilized to illuminate target. Fixed speckles correspond to fixed noise distribution. Consequently, we propose a DnCNN-based GI denoising method using fixed speckles to train and illuminate target in this paper.
A sequence of fixed speckles are utilized to illuminate the samplings in training set. Then, we input the noisy image into the networks, and the output of the networks is predicted residual image.  In this way, a mapping between speckles and residual is formed. We train a denoising networks that the noise is corresponding to a sequence of fixed speckles. Then, the high-resolution target image is obtained by subtracting the predicted residual image from the noisy image. The purpose of this method is to obtain a better target image when using this sequence of fixed speckles to illuminate any target. The loss function of the proposed method can be expressed as
\begin{equation}
{\ell _{GI}}\left( \Theta  \right) = \frac{1}{{2N}}\sum\limits_{s = 1}^N {\left\| {\Re \left( {{y_s};\Theta } \right) - \left( {{y_s} - {x_s}} \right)} \right\|_F^2},
\end{equation}
where $\Re \left( {} \right)$ denotes the residual mapping between speckles and residual, ${x_s}$ is the original image and ${y_s}$ is the noisy image. After training, the sequence of speckles used in the training are utilized to illuminate target, and noisy image input the trained network to predict residual image. Finally, the improved target image is obtained by subtracting the residual image from the noisy image.

\section{The Proposed Method}
\subsection{Principal}
\begin{figure*}[!t]
\centering
\includegraphics[width = 13 cm]{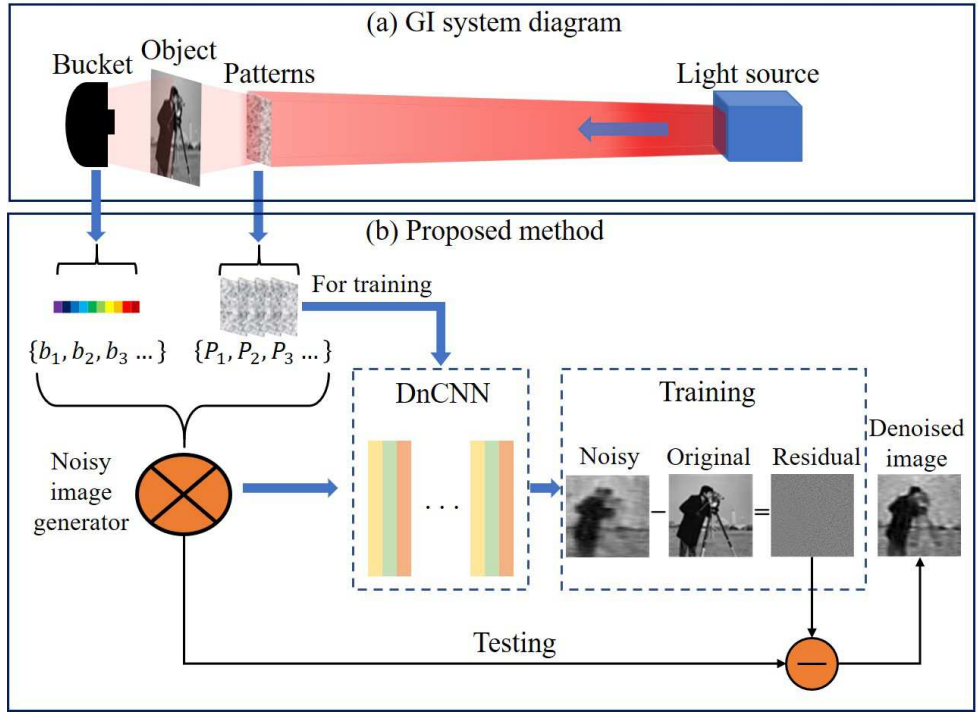}
\caption{Schematic diagram of the proposed method.}
\label{architecture}
\end{figure*}

The image can be recovered from the correlation between the patterns and the bucket signal:
\begin{equation}
G^{(2)}=\mathbf{P}^T\cdot\mathbf{B}=\mathbf{P}^T \mathbf{P}\cdot\mathbf{O},
\end{equation}
$\mathbf{P}^T \mathbf{P}$ plots the convariance map of the two columns of $\mathbf{P}$.  If $\mathbf{P}^T \mathbf{P}=I$ (an identity matrix), $G^{(2)}(\mathbf{r})=\mathbf{O}(r)$. This condition is satisfied, when $\mathbf{P}$ is an orthogonal matrix (such as Hadamard). When $P_j(\mathbf{r})$ is a random pattern, how much $\mathbf{P}^T \mathbf{P}$ is close to $I$ depends on the linear independency of two different coloums of $\mathbf{P}$. Meanwhile, $M$ should be large enough to satisfy the statistic condition. In practics, $M$ can not be set a very large number, otherwise the imaging speed would be highly slowed down. On the other hand, an insufficient large $M$ will worsen the image reconstruction. Let's factorize the correlation into two terms:
\begin{equation}
G^{(2)}=\mathbf{O}+\left[\mathbf{P}^T \mathbf{P}-I\right]\cdot\mathbf{O},
\end{equation}
The second term on the right hand side is called residue in GI, and we can find that speckle is directly related to residual.
$G^{(2)}$ is termed as a noisy image contaminated by the residual image.
Fig.~\ref{architecture} shows the architecture of the proposed method.

\subsection{Architecture}
\begin{figure}[h]
\centering
\includegraphics[width = 8.5 cm]{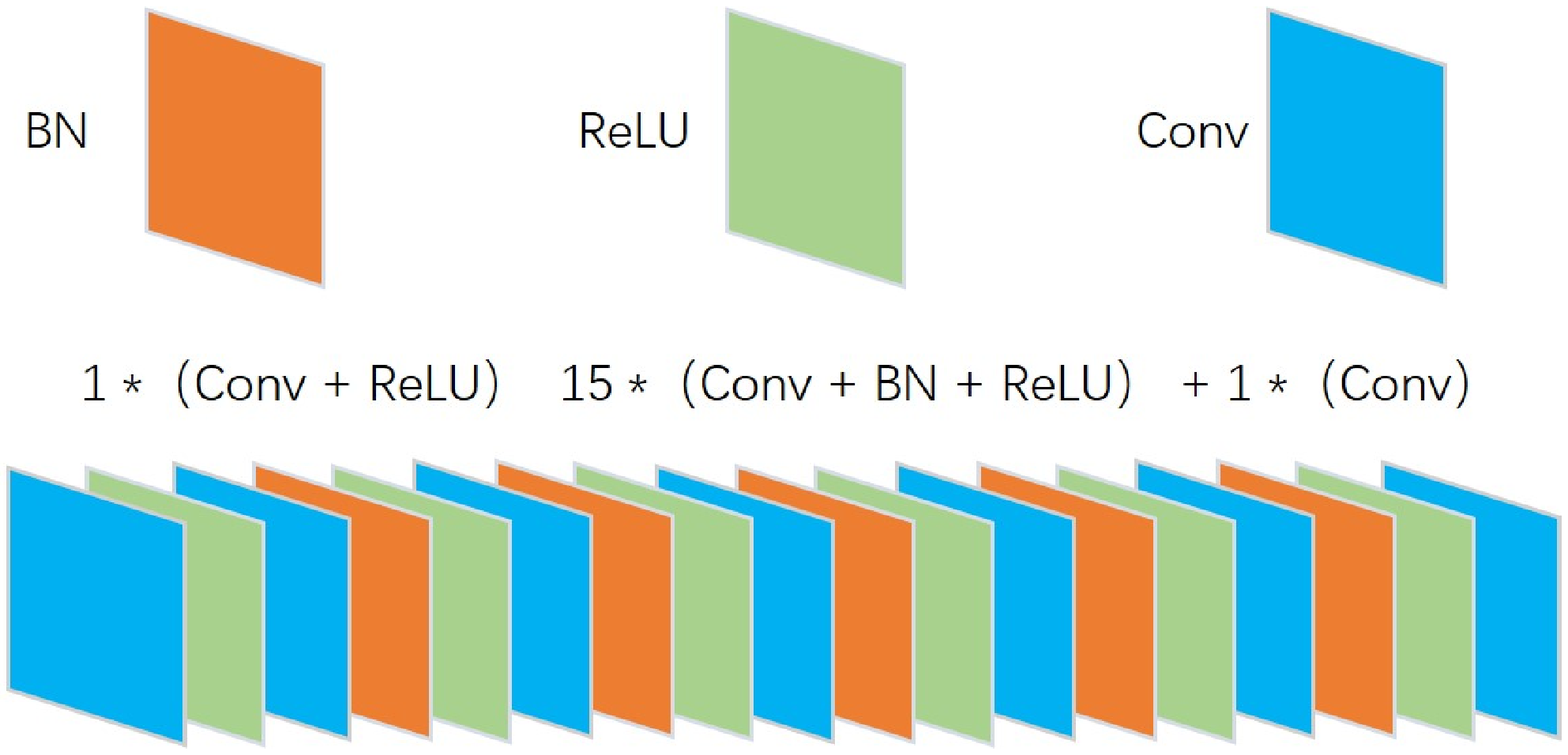}
\caption{Architecture of DnCNN in our work.}
\label{netarchitectures}
\end{figure}

The architecture of the DnCNN in our work is shown in Fig.~\ref{netarchitectures}.

The network has a total of 17 layers. The first layer of the network is composed of a convolutional layer and a ReLU layer, the 2nd to 16th layers are both composed of convolutional layer, ReLU layer and BN layer, and the last layer is a convolutional layer. The padding of all convolutional layer is 1. Because we use single-channel grayscale images for training and testing, the convolutional layer in 1st layer contains 64 filters with size 1*3*3. The 2nd to 16th layers contain 64 filters with size 64*3*3. To ensure the number of input and output channel is 1, the last convolutional layer contains 1 filter with size 64*3*3. Simultaneously, to make the network input and output have the same size, we did not use the Pooling layer in our network.

\section{Experimental Results}
\subsection{Training Settings}
We use 390 images in~\cite{zhang2017beyond} of size 64*64 for training, and set the patch size as 64*64. We train 500 epochs, 1000 epochs, 1500 epochs and 2000 epochs for the DnCNN models, respectively. Before training, we use a sequence of fixed random speckles to sampling every element in training set, and the target image with noise is obtained by GI method. In this way, we prepared two data sets. One is obtained by sampling the data in training set with 2048 speckles, which is defined as sampling rate = 0.5, and the other is obtained by sampling the data in training set with 4096 speckles, which is defined as sampling rate = 1. Then, we use these two data sets to train two DnCNN models.
The equipment used in our experiment is a desktop computer, the CPU is Intel-i7-9750H-2.6GHz, and the graphics is NVIDIA-GeForce-GTX-1660Ti.

\subsection{Results on Set10, Set12 and Set68}
\begin{figure}[h]
\centering
\includegraphics[width = 8.5 cm]{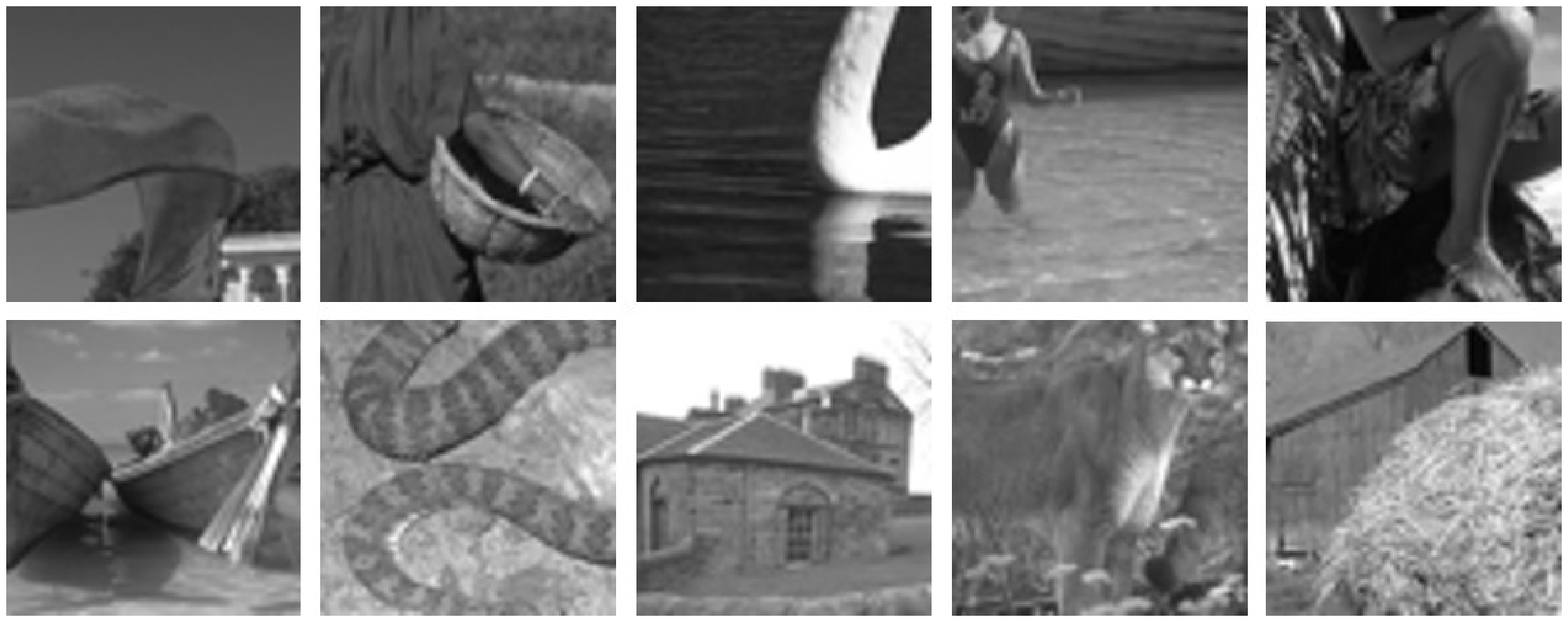}
\caption{Images in Set10.}
\label{set10}
\end{figure}

First, we use the remaining 10 images in~\cite{zhang2017beyond} (defined as Set10 and shown in Fig.~\ref{set10}) to test the proposed method under two different sampling rates, and compare with BC and CS through PSNR/SSIM.

It can be seen from the data listed in Table \uppercase\expandafter{\romannumeral1} that the proposed method can achieve the best PSNR results at both sampling rates. On average, at sampling rate = 0.5, when epoch reaches 1000, the proposed method is 12.81 dB higher than BC algorithm and 0.74 dB higher than CS algorithm. At sampling rate = 1, when epoch reaches 2000, the proposed method is 15.27 dB higher than BC algorithm and 0.87 dB higher than CS algorithm (The CS algorithm used in this paper is orthogonal-matching-pursuit, OMP). Table \uppercase\expandafter{\romannumeral1} shows the PSNR value of each image in Set10.

\begin{table*}[!t]
\caption{Comparison Results of PNSR on Set10}
\label{table_example}
\centering
\begin{tabular}{|c|c|c|c|c|c|c|c|c|c|c|c|}
\hline
Set10 & I1                    & I2                    & I3                    & I4                    & I5                    & I6                    & I7                    & I8                    & I9                    & I10                   & AVE                   \\ \hline
\multicolumn{12}{|c|}{Sampling Rate = 0.5}                                                                                                                                                                                                                                    \\ \hline
BC    & 11.05                 & 10.97                 & 8.82                  & 10.03                 & 9.96                  & 10.04                 & 9.66                  & 7.96                  & 10.51                 & 9.36                  & 9.84                  \\ \hline
CS    & 26.09                 & 23.12                 & 19.10                 & 24.87                 & 18.24                 & 21.43                 & 22.14                 & 20.86                 & 22.01                 & 21.15                 & 21.91                 \\ \hline
\multicolumn{12}{|c|}{Proposed Method}                                                                                                                                                                                                                                        \\ \hline
Epoch & \multicolumn{1}{l|}{} & \multicolumn{1}{l|}{} & \multicolumn{1}{l|}{} & \multicolumn{1}{l|}{} & \multicolumn{1}{l|}{} & \multicolumn{1}{l|}{} & \multicolumn{1}{l|}{} & \multicolumn{1}{l|}{} & \multicolumn{1}{l|}{} & \multicolumn{1}{l|}{} & \multicolumn{1}{l|}{} \\ \hline
500   & 26.49                 & 23.71                 & 20.8                  & 25.55                 & 19.06                 & 21.83                 & 22.59                 & 21.77                 & 22.45                 & 21.83                 & 22.61                 \\ \hline
1000  & \textbf{26.54}        & \textbf{23.88}        & \textbf{20.91}        & \textbf{25.57}        & 18.92                 & 21.74                 & \textbf{22.67}        & 22.13                 & 22.32                 & 21.8                  & \textbf{22.65}        \\ \hline
1500  & 26.26                 & 23.73                 & 20.83                 & 25.41                 & 18.49                 & 21.74                 & 22.6                  & \textbf{22.2}         & \textbf{22.29}        & 21.75                 & 22.53                 \\ \hline
2000  & 26.38                 & 23.6                  & 20.49                 & 25.46                 & \textbf{18.99}        & \textbf{21.81}        & 22.59                 & 21.68                 & 22.44                 & \textbf{21.82}        & 22.53                 \\ \hline
\multicolumn{12}{|l|}{}                                                                                                                                                                                                                                                       \\ \hline
\multicolumn{12}{|c|}{Sampling Rate = 1}                                                                                                                                                                                                                                      \\ \hline
BC    & 11.83                 & 12.21                 & 9.86                  & 10.35                 & 11.41                 & 10.4                  & 9.64                  & 8.04                  & 10.76                 & 9.46                  & 10.40                 \\ \hline
CS    & 28.8                  & 26.16                 & 23.52                 & 27.34                 & 20.35                 & 24.21                 & 24.7                  & 24.44                 & 25.06                 & 23.44                 & 24.8                  \\ \hline
\multicolumn{12}{|c|}{Proposed Method}                                                                                                                                                                                                                                        \\ \hline
Epoch &                       &                       &                       &                       &                       &                       &                       &                       &                       &                       &                       \\ \hline
500   & 29.03                 & 26.58                 & 25.37                 & 27.75                 & 20.99                 & 24.56                 & 24.99                 & 24.52                 & 25.23                 & 23.78                 & 25.28                 \\ \hline
1000  & \textbf{29.22}        & 26.76                 & 27.15                 & \textbf{27.93}        & 21.28                 & \textbf{24.75}        & \textbf{25.04}        & 25.34                 & \textbf{25.28}        & \textbf{23.95}        & \textbf{25.67}        \\ \hline
1500  & 29.06                 & \textbf{26.86}        & \textbf{27.19}        & 27.75                 & \textbf{21.38}        & 24.57                 & 24.9                  & \textbf{25.41}        & 25.23                 & 23.93                 & 25.63                 \\ \hline
2000  & 28.7                  & 26.15                 & 23.83                 & 27.35                 & 20.45                 & 24.33                 & 24.81                 & 24.43                 & 25.1                  & 23.58                 & 24.87                 \\ \hline
\end{tabular}
\end{table*}

In addition, we test the proposed method on the other two larger datasets contain 12 and 68 images under two sampling rates, defined as ExtSet12 and ExtSet68. We test each image in the two datasets and compare the average PSNR/SSIM. We listed the average values calculated from each element in the datasets in Table \uppercase\expandafter{\romannumeral2}. The testing results of the Set12 and Set68 show that the proposed method achieves the best performances in both PSNR and SSIM. Compared to BC and CS algorithm in ExtSet12, the proposed method has a PSNR gain of about 10.17 dB and 0.65 dB at sampling rate = 0.5 and a PSNR gain of about 12.98 dB and 0.67 dB at sampling rate = 1. Similarly, for ExtSet68, the proposed method has 12.06 dB and 0.68 dB improvement at sampling rate = 0.5 and 14.64 dB and 0.75 dB improvement at sampling rate = 1.

\begin{table*}[!t]
\caption{Comparison Results of PNSR and SSIM on ExtSet12 and ExtSet68}
\label{table_example}
\centering
\begin{tabular}{|c|c|c|c|c|c|c|}
\hline
\multicolumn{7}{|l|}{ExtSet12}                                                    \\ \hline
\multicolumn{7}{|c|}{Sampling Rate = 0.5}                                         \\ \hline
Method & BC    & CS    & Epoch=500 & Epoch=1000     & Epoch=1500     & Epoch=2000 \\ \hline
PSNR   & 9.24  & 18.76 & 19.28     & \textbf{19.41} & 19.40          & 19.18      \\ \hline
SSIM   & 0.03  & 0.45  & 0.48      & 0.48           & \textbf{0.49}  & 0.48       \\ \hline
\multicolumn{7}{|c|}{Sampling Rate = 1}                                           \\ \hline
Method & BC    & CS    & Epoch=500 & Epoch=1000     & Epoch=1500     & Epoch=2000 \\ \hline
PSNR   & 9.55  & 21.86 & 21.95     & 22.24          & \textbf{22.53} & 22.46      \\ \hline
SSIM   & 0.06  & 0.68  & 0.68      & 0.7            & \textbf{0.72}  & 0.72       \\ \hline
\multicolumn{7}{|l|}{}                                                            \\ \hline
\multicolumn{7}{|l|}{ExtSet68}                                                    \\ \hline
\multicolumn{7}{|c|}{Sampling Rate = 0.5}                                         \\ \hline
Method & BC    & CS    & Epoch=500 & Epoch=1000     & Epoch=1500     & Epoch=2000 \\ \hline
PSNR   & 9.52  & 20.9  & 21.45     & 21.56          & \textbf{21.58} & 21.41      \\ \hline
SSIM   & 0.03  & 0.51  & 0.54      & 0.54           & \textbf{0.55}  & 0.54       \\ \hline
\multicolumn{7}{|c|}{Sampling Rate = 1}                                           \\ \hline
Method & BC    & CS    & Epoch=500 & Epoch=1000     & Epoch=1500     & Epoch=2000 \\ \hline
PSNR   & 10.03 & 23.92 & 23.97     & 24.36          & \textbf{24.67} & 24.6       \\ \hline
SSIM   & 0.07  & 0.71  & 0.71      & 0.73           & \textbf{0.74}  & 0.74       \\ \hline
\end{tabular}
\end{table*}

\subsection{Comparison With DnCNN}
\begin{figure}[!t]
\centering
\includegraphics[width = 8.5 cm]{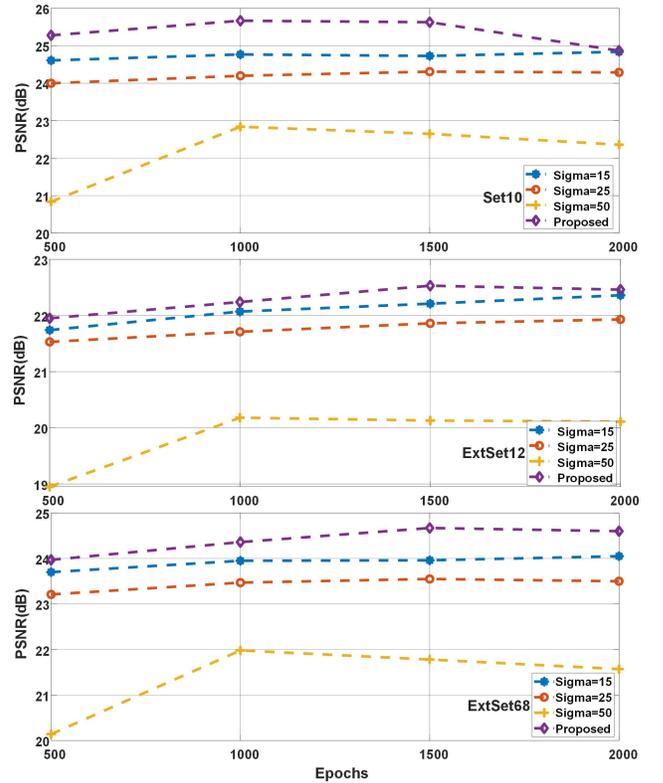}
\caption{Comparison results with DnCNN.}
\label{DNCNN}
\end{figure}

Then, we compare the proposed method with the conventional DnCNN on Set10, ExtSet12 and ExtSet68 at sampling rate = 1. The DnCNN in our comparison testing is trained by the 390 images that used in the training network of the proposed method. We consider three noise levels for DnCNN, i.e., $\sigma$ = 15, 25 and 50. Fig.~\ref{DNCNN} shows the trend chart of PSNR comparison results.
We can find that the performance of the proposed method on each data set is better than DnCNN trained with the same training set. Table \uppercase\expandafter{\romannumeral3} shows the specific values. Compared with the PSNR of the proposed method, we can find that the best PSNR performance of DnCNN on Set10, ExtSet12 and ExtSet68 is 24.84 dB, 22.36 dB and 24.05 dB respectively, which is less than the proposed method 0.83 dB, 0.17 dB and 0.62 dB, respectively.

Noteworthy, the two subsections above (subsection B and C) show that the lower PSNR of the network input, the greater the improvement of the proposed method. The higher PSNR of the input, the limited room for the improvement.

\begin{table*}[!t]
\caption{Comparison Results With DnCNN}
\label{table_example}
\centering
\begin{tabular}{|c|c|c|c|c|}
\hline
\multicolumn{5}{|l|}{Set10}                                   \\ \hline
\multicolumn{5}{|c|}{Sampling Rate = 1}                       \\ \hline
Method & Epoch=500 & Epoch=1000 & Epoch=1500 & Epoch=2000     \\ \hline
s=15   & 24.61     & 24.77      & 24.73      & \textbf{24.84} \\ \hline
s=25   & 24        & 24.2       & 24.31      & 24.29          \\ \hline
s=50   & 20.84     & 22.84      & 22.65      & 22.36          \\ \hline
\multicolumn{5}{|l|}{}                                        \\ \hline
\multicolumn{5}{|l|}{ExtSet12}                                \\ \hline
\multicolumn{5}{|c|}{Sampling Rate = 1}                       \\ \hline
Method & Epoch=500 & Epoch=1000 & Epoch=1500 & Epoch=2000     \\ \hline
s=15   & 21.74     & 22.07      & 22.21      & \textbf{22.36} \\ \hline
s=25   & 21.53     & 21.71      & 21.86      & 21.93          \\ \hline
s=50   & 18.95     & 20.18      & 20.13      & 20.11          \\ \hline
\multicolumn{5}{|l|}{}                                        \\ \hline
\multicolumn{5}{|l|}{ExtSet68}                                \\ \hline
\multicolumn{5}{|c|}{Sampling Rate = 1}                       \\ \hline
Method & Epoch=500 & Epoch=1000 & Epoch=1500 & Epoch=2000     \\ \hline
s=15   & 23.7      & 23.95      & 23.96      & \textbf{24.05} \\ \hline
s=25   & 23.21     & 23.47      & 23.55      & 23.5           \\ \hline
s=50   & 20.14     & 21.98      & 21.78      & 21.57          \\ \hline
\end{tabular}
\end{table*}

\subsection{GI Experimental}
\begin{figure}[h]
\centering
\includegraphics[width = 6 cm]{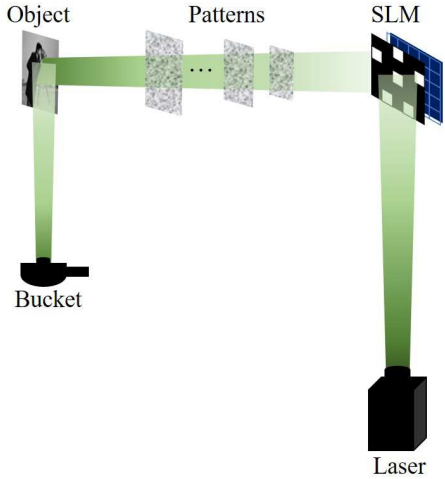}
\caption{Schematic diagram of GI experiment.}
\label{SLM}
\end{figure}

Moreover, we build GI experimental platform to verify the proposed method. The schematic diagram is shown in Fig.~\ref{SLM}.

A expanded laser beam is incident onto a SLM, which modulates the phase of the light field on each pixies.
The SLM is covered by a mask with randomly distributed pinholes.
Only the fields within the pinholes can have the phase modulation.
The size of each pinhole is 0.2 mm.
The number of the pinholes is 24, which are distributed within a circle of 5 mm diameter.
Thus, all above devices construct a light source that has 24 sub-fields.
By manipulating the phases of the sub-fileds, the interference between them generates a variety of light patterns on a distant plane.
However, the number of the sub-fields is insufficient, and the generated patterns are not sufficiently linear independent, which will indue repetitive visual artifacts in the reconstruction image.
The proposed method is able to remove such artifacts and retrieve a desirable image.

\begin{figure*}[!t]
\centering
\includegraphics[width = 18 cm]{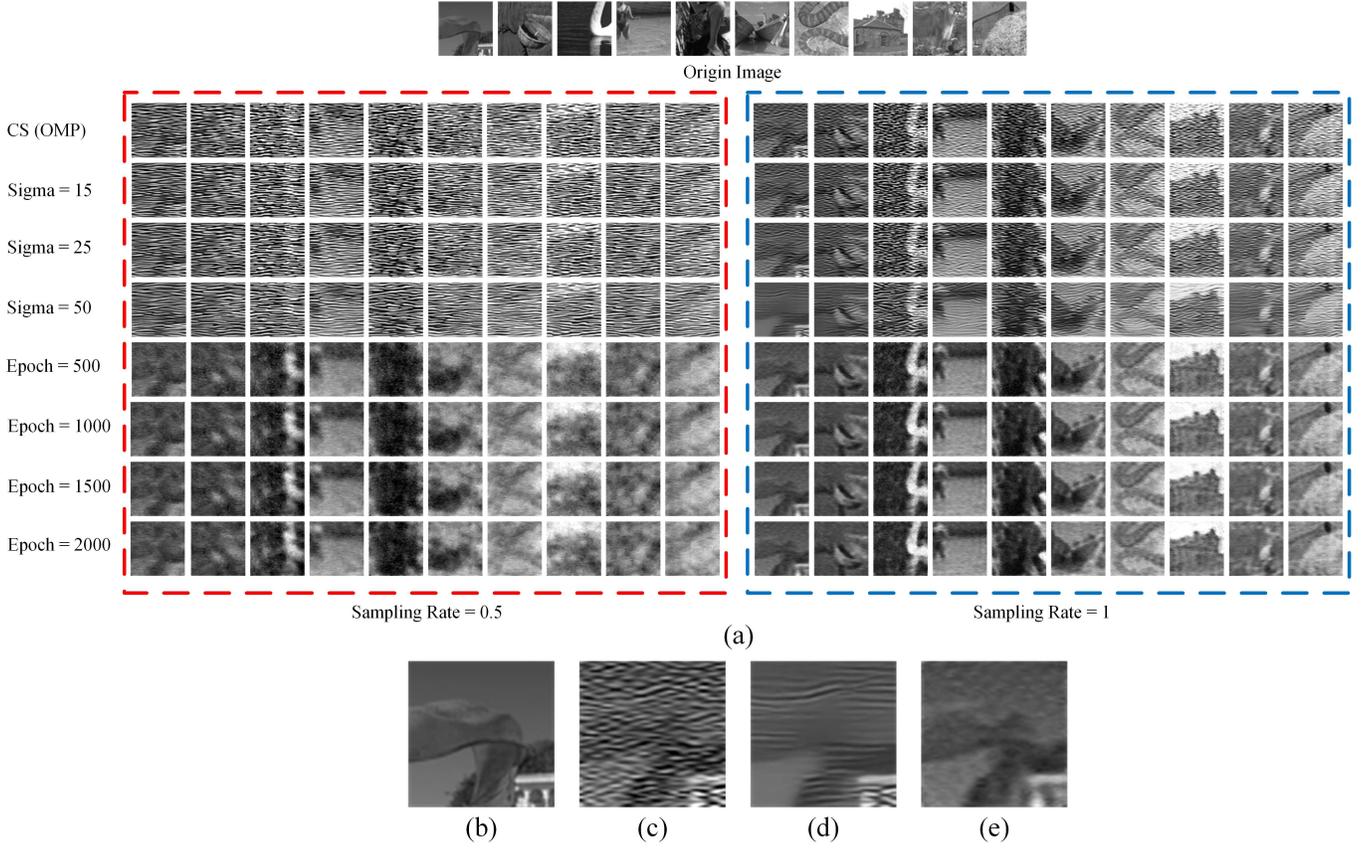}
\caption{Experimental results on Set10. (a) comparison results under different conditions. (b) origin image of the first picture in Set10. (c) Result of CS method at sampling rate = 1. (d) Result of DnCNN with $\sigma$ = 50 at sampling rate = 1. (e) Result of the proposed method with epoch =1500 at sampling rate = 1.}
\label{sanbanduibi}
\end{figure*}

We train a DnCNN using the speckles sequence in the above experiment and test it on Set10. Similarly, we compare the proposed method with CS and DnCNN under two sampling rates. We carry out different $\sigma$ values in DnCNN and different epoch numbers in the proposed method to compare. The results are shown in Fig.~\ref{sanbanduibi}.

Fig.~\ref{sanbanduibi} (a) shows the comparison results under different conditions.
Fig.~\ref{sanbanduibi} (b) shows the origin image of the first picture in Set10.
Fig.~\ref{sanbanduibi} (c) shows the result of CS method (PSNR = 16.25 dB).
Fig.~\ref{sanbanduibi} (d) shows the result of DnCNN (PSNR = 22.83 dB).
Fig.~\ref{sanbanduibi} (e) shows the result of the proposed method (PSNR = 27.22 dB).
From the experimental results, we can see that the proposed method has better effect.
Noteworthy, the performance of the proposed method in physical experiment is better than the simulation data based test.
Detailed comparison results on Set10 are shown in Table \uppercase\expandafter{\romannumeral4}.

\begin{table*}[!t]
\caption{Experimental results on Set10}
\label{table_example}
\centering
\begin{tabular}{|c|c|c|c|c|c|}
\hline
           & I1                  & I2                  & I3                  & I4                  & I5                 \\ \hline
\multicolumn{6}{|c|}{Sampling Rate = 1}                                                                                 \\ \hline
           & PSNR/SSIM           & PSNR/SSIM           & PSNR/SSIM           & PSNR/SSIM           & PSNR/SSIM          \\ \hline
CS         & 16.25/0.15          & 14.7/0.18           & 8.46/0.07           & 13.7/0.11           & 10.57/0.09         \\ \hline
Sigma=15   & 16.98/0.17          & 15.28/0.18          & 8.87/0.08           & 14.19/0.12          & 11.00/0.1          \\ \hline
Sigma=25   & 17.86/0.19          & 15.86/0.19          & 9.02/0.08           & 14.74/0.12          & 11.24/0.1          \\ \hline
Sigma=50   & 22.83/0.48          & 19.46/0.28          & 9.77/0.08           & 17.42/0.2           & 12.3/0.11          \\ \hline
Epoch=500  & 26.42/0.65          & 24.47/0.58          & 18.88/0.38          & 24.9/0.58           & 19.59/0.45         \\ \hline
Epoch=1000 & 26.97/0.7           & 24.86/0.6           & 20.12/0.43          & 25.46/0.62          & 19.92/0.49         \\ \hline
Epoch=1500 & \textbf{27.22/0.73} & \textbf{25.06/0.61} & 20.6/0.45           & \textbf{25.89/0.65} & \textbf{20.17/0.5} \\ \hline
Epoch=2000 & 27.07/0.71          & 24.99/0.61          & \textbf{20.81/0.46} & 25.7/0.63           & 20.14/0.5          \\ \hline
           & I6                  & I7                  & I8                  & I9                  & I10                \\ \hline
           & PSNR/SSIM           & PSNR/SSIM           & PSNR/SSIM           & PSNR/SSIM           & PSNR/SSIM          \\ \hline
CS         & 12.34/0.13          & 12.83/0.16          & 11.56/0.14          & 14.41/0.19          & 12.06/0.14         \\ \hline
Sigma=15   & 12.77/0.13          & 13.28/0.16          & 11.94/0.15          & 14.94/0.2           & 12.5/0.14          \\ \hline
Sigma=25   & 13.2/0.14           & 13.79/0.16          & 12.36/0.15          & 15.51/0.2           & 12.9/0.14          \\ \hline
Sigma=50   & 14.74/0.16          & 15.62/0.18          & 13.67/0.16          & 18.12/0.27          & 14.48/0.16         \\ \hline
Epoch=500  & 22.69/0.6           & 22.8/0.54           & 21.41/0.55          & 23.46/0.6           & 21.78/0.48         \\ \hline
Epoch=1000 & 23.17/0.63          & 23.12/0.56          & 21.96/0.57          & 23.89/0.63          & 22.08/0.5          \\ \hline
Epoch=1500 & \textbf{23.38/0.65} & \textbf{23.4/58}    & 22.66/0.6           & \textbf{24.01/0.64} & 22.28/0.5          \\ \hline
Epoch=2000 & 23.29/0.63          & 23.38/0.57          & \textbf{22.66/0.61} & 23.99/0.64          & \textbf{22.31/0.5} \\ \hline
\end{tabular}
\end{table*}

\subsection{Comparison of Different Speckles Sequences}
\begin{figure}[h]
\centering
\includegraphics[width = 8.5 cm]{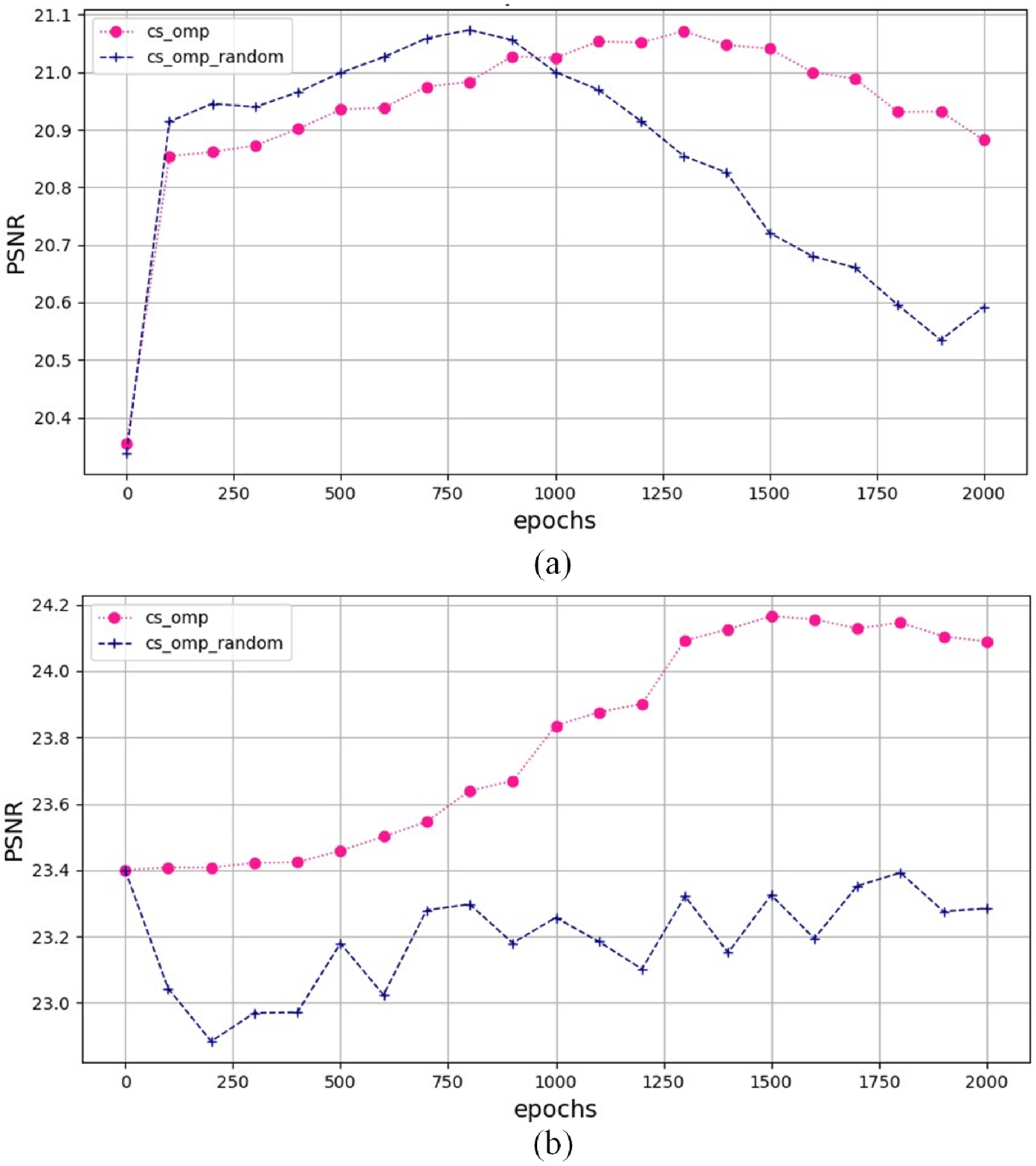}
\caption{PSNR results of different speckles sequences. (a) Sampling rate = 0.5. (b) Sampling rate = 1.}
\label{speckles}
\end{figure}

To test the denoising ability of the proposed method on the corresponding speckles sequence, we use two sets of different speckles sequences to compare. The results are shown in Fig.~\ref{speckles}.

The red curve is the testing result of the DnCNN trained by the speckles sequence corresponding to this curve, and the blue curve is the testing result of another speckles sequence on this network. Fig. 6 shows that no matter which sampling rate, the effect of the proposed will descend when the illumination speckles sequence does not match the training speckles sequence.

\subsection{Data Enhancement}
\begin{figure}[h]
\centering
\includegraphics[width = 8.5 cm]{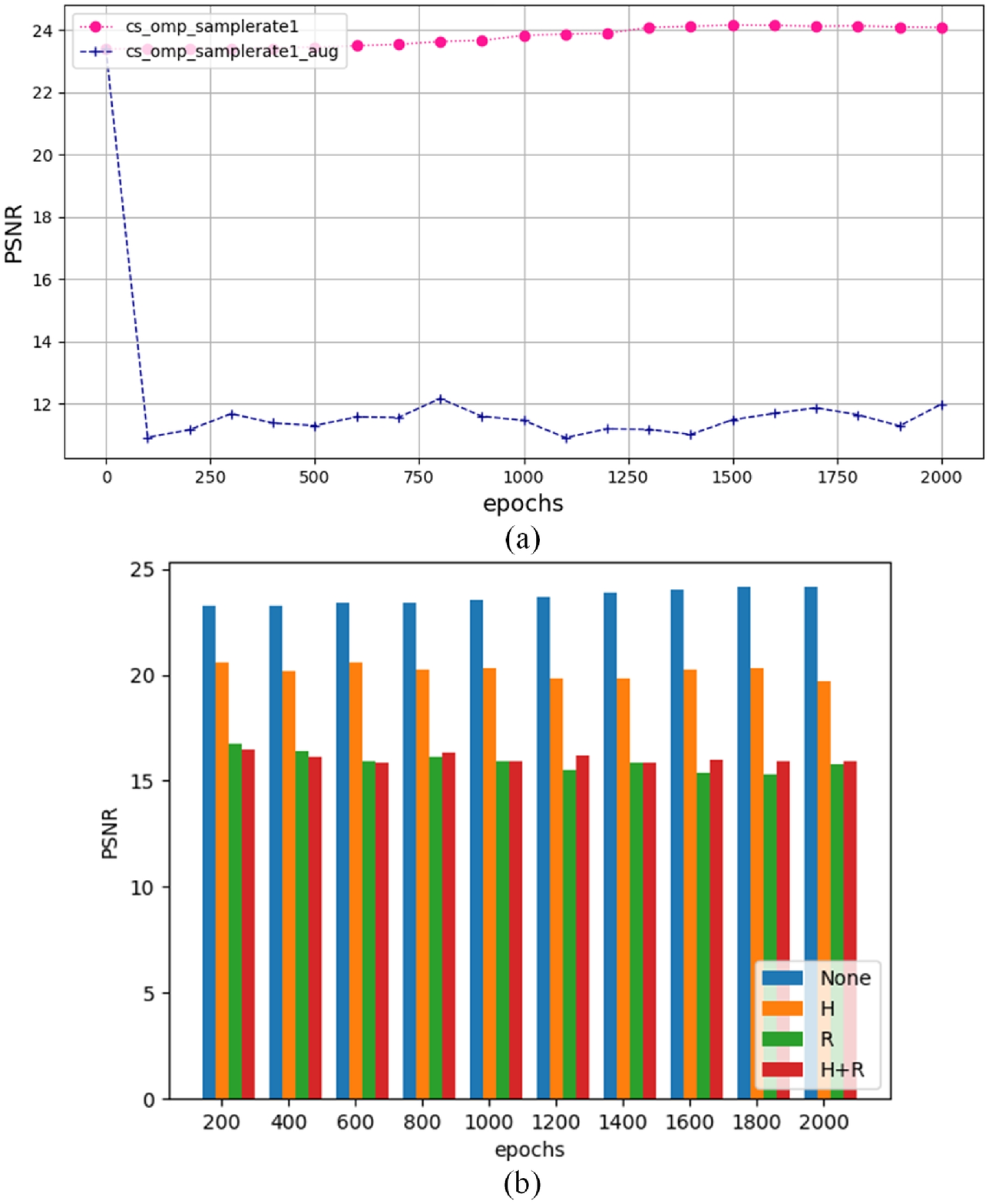}
\caption{Comparison results of data enhancement. (a) Comparison results between the proposed method and conventional DnCNN. (b) Results of four different collocations.}
\label{data}
\end{figure}

In deep learning method, data enhancement can generate new training samples to prevent over fitting. RandomHorizontalFlip and RandomRotation are used as data enhancement strategies in DnCNN. However, we find that using these two data enhancement strategies in the proposed method will affect the performance through experiments. The PSNR decrease to half of that before denoise (shown in Fig.~\ref{data}). We infer that the noise in the image obtained by the fixed speckles sequence is related to the position, the data enhancement of spatial transformation such as RandomHorizontalFlip and RandomRotation will destroy the noise distribution in the image, resulting in the network can not learn the original noise distribution in the image, so the image quality after denoising will not increase but decrease. The comparison results are shown in Fig.~\ref{data}.

We carry out several different sets of contrast experiments including enable RandomHorizontalFlip and RandomRotation (H+R), enable RandomHorizontalFlip (H), enable RandomRotation (R) and close RandomHorizontalFlip and RandomRotation (None).
Fig.~\ref{data} (a) shows the comparison results at sampling rate = 1 between the proposed method and conventional DnCNN that enable RandomHorizontalFlip and RandomRotation. Fig.~\ref{data} (b) shows the results of four different collocations, we can find that for training DnCNN with the proposed method, the effect of closing data enhancement is the best. On the other hand, for conventional DnCNN, the effect of enabling RandomHorizontalFlip and RandomRotation is the best.

\section{Conclusion}
In this paper, we proposed a improved GI method based on DnCNN, which mapping the relationship between speckles and noises in GI.
A set of speckles sequence is employed to train DnCNN based on the mechanism of GI.
In application, we use this speckles sequence in training process to illuminate unknown targets.
Since the noise distribution corresponding to this speckles sequence has been learned, promising performance can be obtained.
Based on three different testing data sets and two sampling rates, extensive experiments have carried out.
Compared with other traditional methods in GI, or DnCNN, the results show that the proposed method has a better performance in PSNR/SSIM.
Especially when the input PSNR is low, the improvement is great.
Moreover, the proposed method can be regarded as a general method for GI, or those imaging methods that using random radiation fields to illuminate target.

\ifCLASSOPTIONcaptionsoff
  \newpage
\fi

\bibliographystyle{IEEEtran}
\bibliography{TIP}

\end{document}